\documentclass[aps, prd, amsmath, floats, floatfix, twocolumn, nofootinbib, 
]{revtex4-1}
\usepackage{graphicx}
\usepackage{color}
\usepackage{latexsym}
\usepackage{float}
\usepackage{amssymb}

\newcommand{\be}{\begin{eqnarray}}
\newcommand{\ee}{\end{eqnarray}}

\newcommand{\dd}{ \mathrm{d}}

\usepackage{comment}

\begin{document}

\title{A note on reflection positivity in nonlocal gravity}

\author{Marios Christodoulou}
\email{christod.marios@gmail.com}
\affiliation{Department of Physics, Southern University of Science and Technology, Shenzhen 518055, China} 

\author{Leonardo Modesto}
\email{lmodesto@sustc.edu.cn} 
\affiliation{Department of Physics, Southern University of Science and Technology, Shenzhen 518055, China}

\date{\today}

\begin{abstract}
Contrary to recent claims in the literature, a simple test for reflection positivite, which we call perturbative reflection positivity in the coincidence limit, is shown to be satisfied for nonlocal field theories. Particular attention is given to weakly nonlocal gravity and gauge theories.  
 
\end{abstract}


\maketitle



{\em Introduction ---}
The quantization of Einstein's theory in the quantum field theory framework runs into a notorious renormalizability issue.  Two set proposals for a modified action for gravity that directly address this problem are currently investigated: weakly nonlocal quantum gravity 
\cite{Krasnikov, kuzmin,tomboulis,modesto,modestoLeslaw} and Lee-Wick quantum gravity \cite{shapiromodesto,LWqg}. 
Prescriptions to ensure perturbative unitarity of Lee-Wick theories have been been proposed in \cite{leewick1,leewick2,CLOP,Piva1,Piva2,Piva3}.
For perturbative unitarity in nonlocal gravity see \cite{tomboulis,Tomboulis:2015gfa, ModestoFabio} and references therein.

 Reflection positivity, one of the Wightman axioms for Minkowski quantum field theory, is an important test for any such theory. Recently, concerns were raised \cite{Asorey2} that weakly non local theories fail to pass a the basic test of perturbative reflection positivity in the coincidence limit. In this note we show that this issue does not arise.

\medskip

{\it The theory ---} 
We begin by recalling the class of gravitational theories investigated in this paper. The action reads \cite{kuzmin,tomboulis,modesto,modestoLeslaw} 
\begin{align} 
S = \frac{2}{\kappa^{2}_D} \! \int \!  d^D x \sqrt{-g} \left[ R + G_{\mu\nu} \gamma(\Box) R^{\mu\nu} + V(\mathcal{R}) \right] \,,
 \label{theory}
\end{align}
where $\kappa_D^2 = 32 \pi G_D$, $G_D$ is Newton's constant in $D$ dimensions, $g$ is the determinant of the metric $g_{\mu\nu}$, and $\Box$ denotes the d'Alembertian operator.
The action consists of three operators, one linear and two quadratic in the Riemann tensor, plus a potential $V(\mathcal{R})$.  The potential is at least cubic in the Riemann tensor, with the notation $\mathcal{R}$ shorthand for any invariant combination of the Ricci scalar, the Ricci tensor and the Riemann tensor, and derivatives thereof.

 The nonlocality of the theory \eqref{theory} is embodied by the inclusion of a function $\gamma(z)$ that is entire in $z$ and is of the form 
\begin{equation}
\gamma(\Box) = \frac{e^{H(\sigma \, \Box)} -1}{\Box} \,,
\end{equation}
with $H(\sigma \, z)$ an entire function in $z$. It is customary to call 
\begin{equation}\label{eq:formFac}
F 
\equiv e^{-H(\sigma \, \Box )} 
\end{equation}
the \emph{form factor}. The form factor includes the dependence on the nonlocality scale $\sigma$, a positive real number. The energy scale at which nonlocal effects become important is of the order $1/\sqrt{\sigma}$. A nonlocal theory for which the form factor is an entire function is called weakly nonlocal.

In order for the counter--terms of nonlocal gravity to be local, the function $H(\sigma \, z )$ seen as a function of a complex variable $z$ must be asymptotically polynomial in a conical region which includes the real axis \cite{kuzmin,tomboulis}. Two popular classes of theories satisfying this property and examined in what follows are given by
\begin{align}
H_K(z) & = \alpha \left[ \log \left(z\right)+\Gamma \left(0,z\right)+\gamma_E \right]\, , \ \textrm{Re} \, z>0\; , \label{HKuzmin}
 \\
H_T( p) & = \frac{1}{2} \left[ \log \left(p^2\right)+\Gamma \left(0,p^2\right)+\gamma_E \right]\, , \  \textrm{Re} \, p^2>0 \, .
\label{HTombo}
\end{align}
Here, $\alpha$ is an integer, $p$ is a polynomial of degree $n$ in the variable $z \equiv \sigma \Box$ and $\gamma_E$ is the Euler-Mascheroni constant. The parameters $\alpha$ and $n$ must satisfy the super--renormalizability conditions
\be
 \alpha > D-1 \; , \quad 
n > D-1 \; , 
\label{eq:superCondT}
\ee
respectively. The inclusion of a non--trivial form factor ($F \neq 1$) in \eqref{theory} modifies the graviton propagator, but does not introduce any extra pole besides the graviton. Concretely, in momentum space and neglecting the gauge dependent terms, the propagator reads \cite{modesto}
\be
G(k) = \frac{e^{-H(\sigma k^2) }}{i (k^2 - i \epsilon) } \left( P^{(2)} - \frac{1}{D-2} P^{(0)} \right)  \, ,
\label{NLP}
\ee
where $P^{(2)}$ and $P^{(0)}$ are the usual spin--two and spin--zero projector operators \cite{modesto}.

The term in parenthesis in \eqref{NLP} is not relevant in what follows and is neglected. That is, it is sufficient to consider a scalar field theory. The results given below extend to gravity by making the tensorial structure explicit, contracting the two-point Green's function with two general conserved energy--momentum tensors, and taking into account the contribution of the Faddeev--Popov ghosts.  

We will also examine form factors that are asymptotically exponential and are of the form
\be
F=e^{-H} = e^{-( - \sigma \Box)^n } \ , \; n \in \mathbb{N}^+ \, .
\label{expFF}
\ee
Such form factors have been studied in the context of string theory and gauge theories, see for instance \cite{collective,Mtheory}.

\medskip

{\em Reflection positivity ---} 
Below we briefly recall the definition of reflection positivity and the necessary condition examined here. We are following \cite{Trinchero} to which we refer the reader for further details. For an analysis of reflection positivity for higher derivative theories see \cite{Arici:2017whq}.

Reflection positivity is one of the axioms laid out by Osterwalder and Schrader in their seminal works   \cite{Osterwalder:1973dx,Osterwalder:1974tc}. It is a necessary and sufficient condition in order for  Euclidean Green's functions to (uniquely) define a Wightman quantum field theory in Minkowski space. A Euclidean quantum field theory is said to satisfy reflection positivity when the following statement holds. For any functional $\mathcal{F}[\phi]$ of the fields whose support includes only points that have positive Euclidean time $\tau >0$, we have  
\begin{equation}
\langle (\Theta\mathcal{F}[\phi])\mathcal{F}[\phi] \rangle \geqslant 0 \, ,   \label{eq:rp}
\end{equation}
where $\Theta\mathcal{F}[\phi]$
denotes complex conjugation and reflection with respect to the $\tau=0$ (hyper) plane. 
To show reflection positivity for a free or interacting theory defined at the perturbative level,
it is sufficient to consider the case when the functional $\mathcal{F}[\phi]$
is linear in $\phi$ and $S[\phi]$ is quadratic in the field and its
derivatives. That is, it is sufficient to consider a field functional
\be \label{eq:fieldFun}
\mathcal{F}[\phi]=\int \dd^{D}x\,f(x)\phi(x),
\ee
where the test function $f(x)$ has support on points for which $\tau>0$. 

Using standard textbook techniques, the expectation value of eq.\! \eqref{eq:rp} for the field functional \eqref{eq:fieldFun} can be written as
\be \label{eq:RP}
&& \hspace{-0.4cm}
\langle (\Theta\mathcal{F}[\phi])\mathcal{F}[\phi]\rangle =\int \dd^{D}x\,\dd^{D}y\,\bar{f}(x)G(\theta x-y)f(y) \, ,
\ee
where $G(x-y)$ denotes the propagator for the operator in $S[\phi]$ (tree--level Schwinger function, two--point Green's function). The bar indicates complex conjugation and the reflection $\theta$ flips the sign of the temporal component of a Euclidean vector $x$,
\begin{equation}
\theta x = \theta(\tau_x,x_1,\ldots,x_{D-1}) = (-\tau_x,x_1,\ldots,x_{D-1}). 
\end{equation}
Then, perturbative reflection positivity is equivalent to demanding that the integral on the right hand side of eq.\! \eqref{eq:RP} is positive for every test function with support on points that have $\tau >0$. 

The test function $f$ can be taken in general to be any tempered distribution in $\mathbb{R}^D$. In this note we restrict to test functions \begin{align} \label{eq:source}
f(x) & =  \delta(x-X).
\end{align}
 Considering only one ``charge'' \cite{Uhlmann1979} at fixed position $X$ is physically equivalent to studying the properties of the propagator in the coincidence limit. Plugging this in \eqref{eq:RP}, gives
\begin{equation}
\langle (\Theta\mathcal{F}[\phi])\mathcal{F}[\phi] \rangle  = G(\theta X - X).
\end{equation}
Thus, a simple necessary (but not sufficient) condition for perturbative reflection positivity is
\begin{align} \label{RPcrite}
G(\theta X -X) \geq 0. 
\end{align}

This is in fact simply the requirement that the propagator $G(Y)$ is positive for an arbitrary Euclidean vector $Y$. This can be seen as follows. Since $\theta X -X = (-2 \tau_X,0,\cdots,0)$, and recalling that $G(Y)$ is really a function only of the norm $\vert Y \vert$ (see eq.\! \eqref{propcoord}\,), we have that
\begin{align}\label{RPs}
 G(\vert \theta X -X \vert)&=  G(\vert (-2 \tau_X,0,\cdots) \vert) \nonumber \\
 &= G(\vert (2 \tau_X,0,\cdots) \vert) \nonumber \\
 & = G(\vert (\tau_{\tilde{X}},0,\cdots) \vert).
\end{align}
for some Euclidean vector $\tilde{X}$. That is,  \eqref{RPcrite} is equivalent to demanding that $G(\tilde{X})$ seen as a function of a  Euclidean vector of the form $\tilde{X} = (\tau_{\tilde{X}},0,0,0)$ is positive. Due to the invariance of the Euclidean inner product and integration measure under rotations (see eq.\! \eqref{Gx}\,), $G(\tilde{X})$ is invariant under arbitrary rotations of $\tilde{X}$. Since there always exists a rotation bringing any vector to the form $\tilde{X}$, \eqref{RPcrite} is equivalent to the requirement that the propagator itself is positive for any vector $Y$
\begin{align} \label{RPcrit2}
G(Y) \geq 0 \quad \forall \; Y  \, .
\end{align}
The above condition and its equivalent form \eqref{RPcrite} are called here perturbative reflection positivity in the coincidence limit.

\medskip

{\em Propagator ---} 
We now simplify the expression for the propagator $G(x)$ for a general form factor $F \equiv \exp - H$ and bring it to a form suitable for the analysis that follows. Similar manipulations can be found in \cite{Efimov} (in russian). Neglecting the tensorial structure as explained above, the Fourier transform of the Euclidean version of \eqref{NLP} reads
\be
G(x) = \int \frac{d^D k}{(2 \pi)^D} \, \frac{F(k^2 \, \sigma)}{k^2} \, e^{ i k \cdot x}   \, .
\label{Gx}
\ee 
Changing from the Cartesian momentum space coordinates $k$ to $D$-dimensional spherical coordinates, and exploiting the invariance of the integration measure and the Euclidean inner product over rotations, the angular coordinates can be integrated out to arrive at
\begin{align} \label{propcoord}
G(\vert x \vert ) = \frac{2^{1-D}}{ \pi^{D/2} } & \; \int_0^\infty \! \dd u \; u^{D-3} \; F(u^2 \, \sigma) \; \nonumber \\
& \ \ {}_0\tilde{F}_1 \left[\frac{D}{2};-\frac{1}{4} u^2 \vert x \vert^2 \right].
\end{align}
Here, we have introduced the radial coordinate in momentum space $u = \sqrt{k^2}$, and $\,_0\tilde{F}_1(a;z) = \,_0{F}_1(a;z)/\Gamma(a)$ is the regularized confluent hypergeometric function.

As an aside, note that the necessary asymptotic behaviour of the form factor can be read from the above expression. Setting $x=0$, and since $\,_0\tilde{F}_1(a;0)=1/\Gamma(a)$, we have
\begin{align} \label{propCoincLim}
G(0) = \frac{2^{1-D}}{ \pi^{D/2} \, \Gamma(D/2) } & \; \int_0^\infty \! \dd u \; u^{D-3} \; F(u^2 \, \sigma) \, .
\end{align}
Setting $F=1$, corresponding to the Hilbert-Einstein action, gives an infinite result as it should. Assuming a form factor with the asymptotic behaviour 
\begin{equation}
F(u^2 \, \sigma ) = O(1/ u^{2 A})  \, , \quad A > 0 \,  ,
\end{equation}
and considering only the upper integration limit, we find that $G(0)$ is finite only if 
\begin{equation}
A > \frac{D-2}{2} \, ,
\end{equation}
which is satisfied for the form factors (\ref{HKuzmin}) and (\ref{HTombo}), for the parameters as in 
eq.\,\eqref{eq:superCondT} .
\bigskip

 Returning to reflection positivity, from the form \eqref{propcoord} of the propagator we can prove the following statement.
{\em For any form factor $F(u)$ that is a bounded positive monotonically decreasing function of the non-negative variable $u$, the propagator $G(Y)$ is positive for any Euclidean vector $Y$.}
This is our main technical result. It follows from the following integral inequality \cite{Inequality}. For any bounded positive real function $F(u)$ that is monotonically decreasing on the positive real axis (that is, $F' < 0$ and $ 0<F<\infty $ for $u \in (0,\infty)$\,), and for any $J(u)$ that satisfies
\begin{equation}
\int_0^u \dd t \; J(t) >  0 \ , \ \;  \forall u \in (0,\infty),
\end{equation}
we have that
\begin{equation}
\int^\infty_0 \dd u \; F(u) J(u) >  0.
\end{equation}
Proving the above inequality is an elementary exercise in calculus. It suffices to use partial integration to arrive at 
\begin{align}
\int_0^\infty\! F(u) J(u) \dd u & = F(u)  \int_0^u J(t) \dd t \, \bigg\vert_0^\infty \nonumber \\ &- \int_0^\infty F'(u) \left[ \int_0^u J(t) \dd t \right] \dd u \,  ,
\end{align}
and notice that the right hand side is the sum of two strictly positive terms by assumption. 

We now restrict to four dimensions. Setting $D=4$ and using the identity
\begin{equation} \label{eq:besselIdent}
\frac{ 2\,  J_1 \! \left(\, u \vert x \vert \, \right)}{u \vert x \vert}=
   {}_0\tilde{F}_1\left[2;-\frac{1}{4} u^2 \vert x \vert^2\right],
\end{equation}
the propagator simplifies to 
\begin{equation}
G(\vert x \vert ) = \frac{1}{4 \pi^2 \, \vert x \vert} \, \int_0^\infty \! \dd u  \; F(u^2 \sigma) \;  J_1 \! \left(\, u \vert x \vert \, \right),
\label{propCoordD4}
\end{equation}
where $J_n$ are the Bessel functions of the first kind. Thus, we need only show that 
\begin{equation} \label{eq:statement}
\int_0^u \dd t \; J_1 \! \left(\, u \vert x \vert \, \right) >  0 \ , \ \;  \forall u, \vert x\vert \in (0,\infty).
\end{equation}
Indeed, we have that 
\begin{equation}\label{identity} 
\int_0^u \dd t \; J_1 \! \left(\, t \vert x \vert \, \right) = \frac{1- J_0(u \vert x \vert )}{\vert x \vert} > 0
 \end{equation}
because $J_0(y )$ is always less than unit for $y \in (0,\infty)$ and thus, the inequality \eqref{eq:statement} holds. Since the form factor is a positive function by its definition \eqref{eq:formFac}, we conclude that the criterion \eqref{RPcrit2} is always satisfied for any form factor $F(u^2 \sigma)$ that is a monotonically decreasing function of $u$. That is, a sufficient condition for perturbative reflection positivity in the coincidence limit is simply that the form factor is a monotonically decreasing function
\begin{equation} \label{eq:suff}
F' < 0 \, .
\end{equation}

All form factors of the form \eqref{expFF} satisfy the above. This result is in contrast with the concerns raised in \cite{Asorey2}, where such form factors and the test \eqref{RPcrite} were considered. For form factors as in eq. \! \eqref{HTombo} we have that 
\begin{equation}
\frac{\dd F}{\dd u} = - \frac{\dd p}{\dd u} \, \frac{\dd H}{\dd p} \; F.
\end{equation}
Since
\begin{equation}
\frac{\dd H}{\dd p} = \frac{\alpha}{p} \left(1- e^{-p^2} \right),
\end{equation}
the inequality \eqref{eq:suff}, and thus also \eqref{RPcrit2}, is automatically satisfied whenever $p$ is a monomial, which includes the case of eq.\,\eqref{HKuzmin}. Then, \eqref{RPcrit2} and by extension \eqref{RPcrite} is shown to hold for these cases as well.

\medskip
Above, we have given a proof for the positivity of the propagators for a wide class of nonlocal theories, and explained that perturbative reflection positivity in the coincidence limit follows. We stress however that as mentioned above, the criterion \eqref{eq:suff} provides a sufficient condition, not a necessary condition, and its failure does not necessarily imply that \eqref{RPcrit2} is not satisfied. Furthermore, to show the inequality \eqref{eq:statement} we restricted to four dimensions. For higher dimensions, hypergeometric functions must be considered as in eq.\! \eqref{propcoord} and an identity that simplifies matters as in eq.\! \eqref{identity} is, to our knowledge, not available,  making analytic manipulations difficult. 

\begin{figure}[H]
  \centering
\includegraphics[scale=0.18]{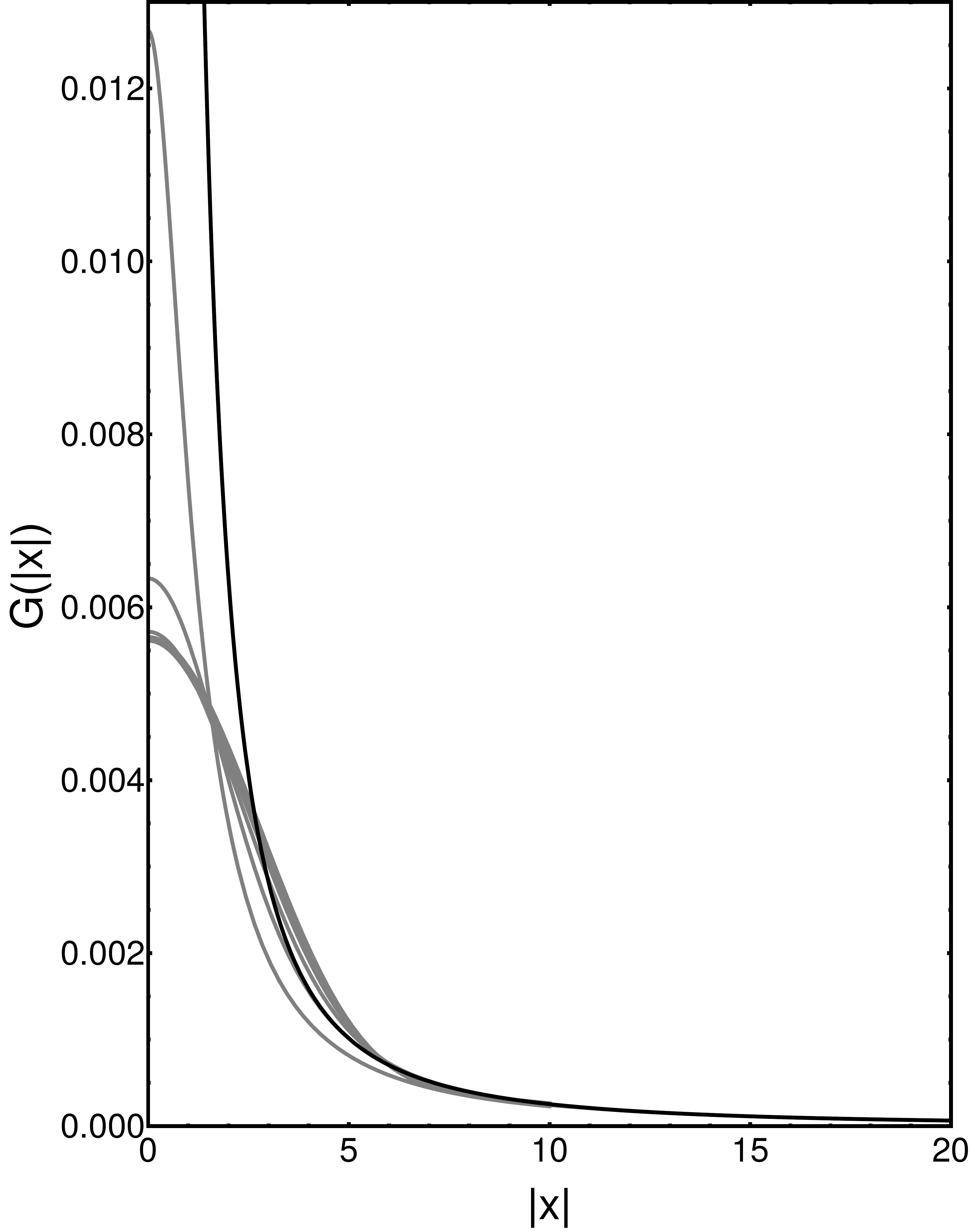}
\caption{The propagators for the form factor of eq. \eqref{expFF} for $D=4$ and $n=1/2,1,3/2,2,5/2,3$ (gray). The propagator for General Relativity is drawn for comparison (black). The propagators for the non-local theory are positive and well behaved in the ultraviolet, including the coincidence limit $\vert x \vert=0$. The infrared behaviour is identical as that of General Relativity. We note that as $n$ increases, $G(0)$ converges to the value $\sim 6 \times 10^{-3}$. The nonlocality scale $\sigma$ has been set to unit here. \vspace*{1cm}}
\label{fig:d4E}
\end{figure}

\begin{figure}[H]
\centering
\includegraphics[scale=0.18]{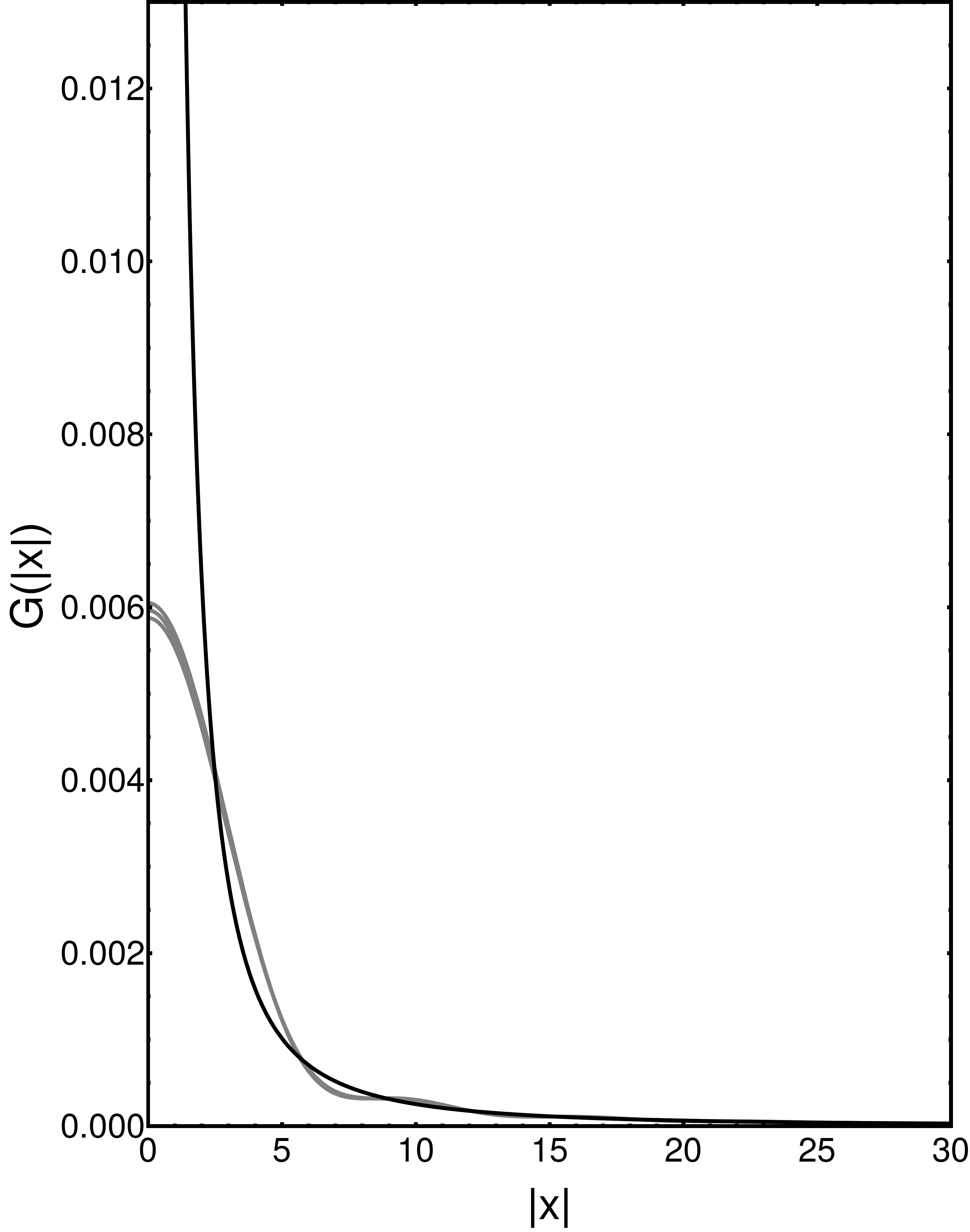}
\caption{The propagators for the form factor of eq. \eqref{HTombo} for $D=4$, and $p=z^k$ with $k=3,6,8, 11$ (gray). The behaviour is qualitatively similar as in Fig. \ref{fig:d4E}. }
\label{fig:d4NL}
\end{figure}

\begin{figure}[H]
\centering
\includegraphics[scale=0.18]{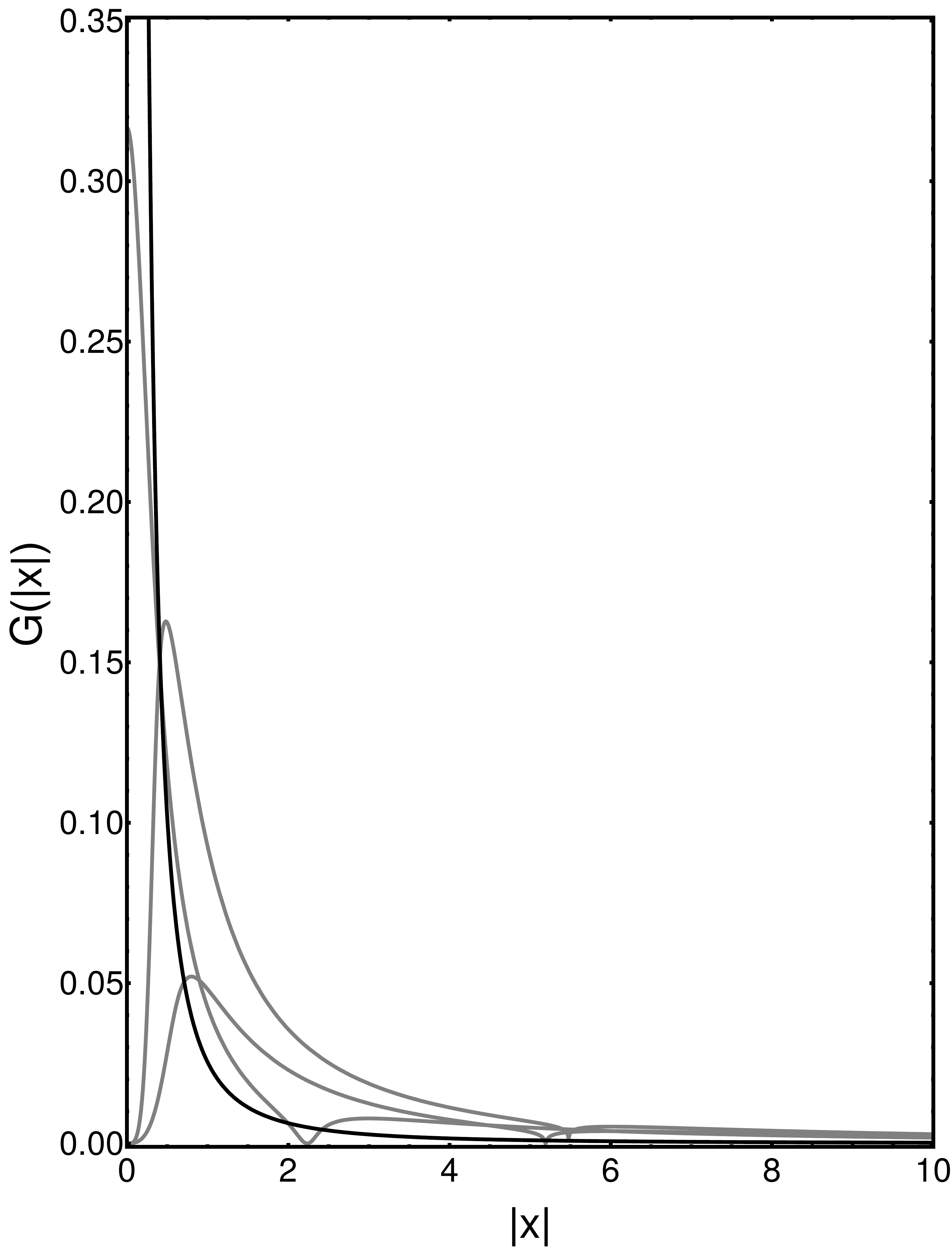}
\caption{The propagators for the form factor of eq. \eqref{HTombo} for $D=4$, for three examples of polynomials that are relatively badly behaved and violate the sufficient condition \eqref{eq:suff}: $p=u^5-30u^3$, $p=u^4-u^2-5u^3$, $p=u^3-5u$. Numerical studies verify that for arbitrary polynomials the propagators never become negative. }
\label{fig:d4NLVio}
\end{figure}

\begin{figure}[H]
\vspace*{2cm}
\centering
\includegraphics[scale=0.186]{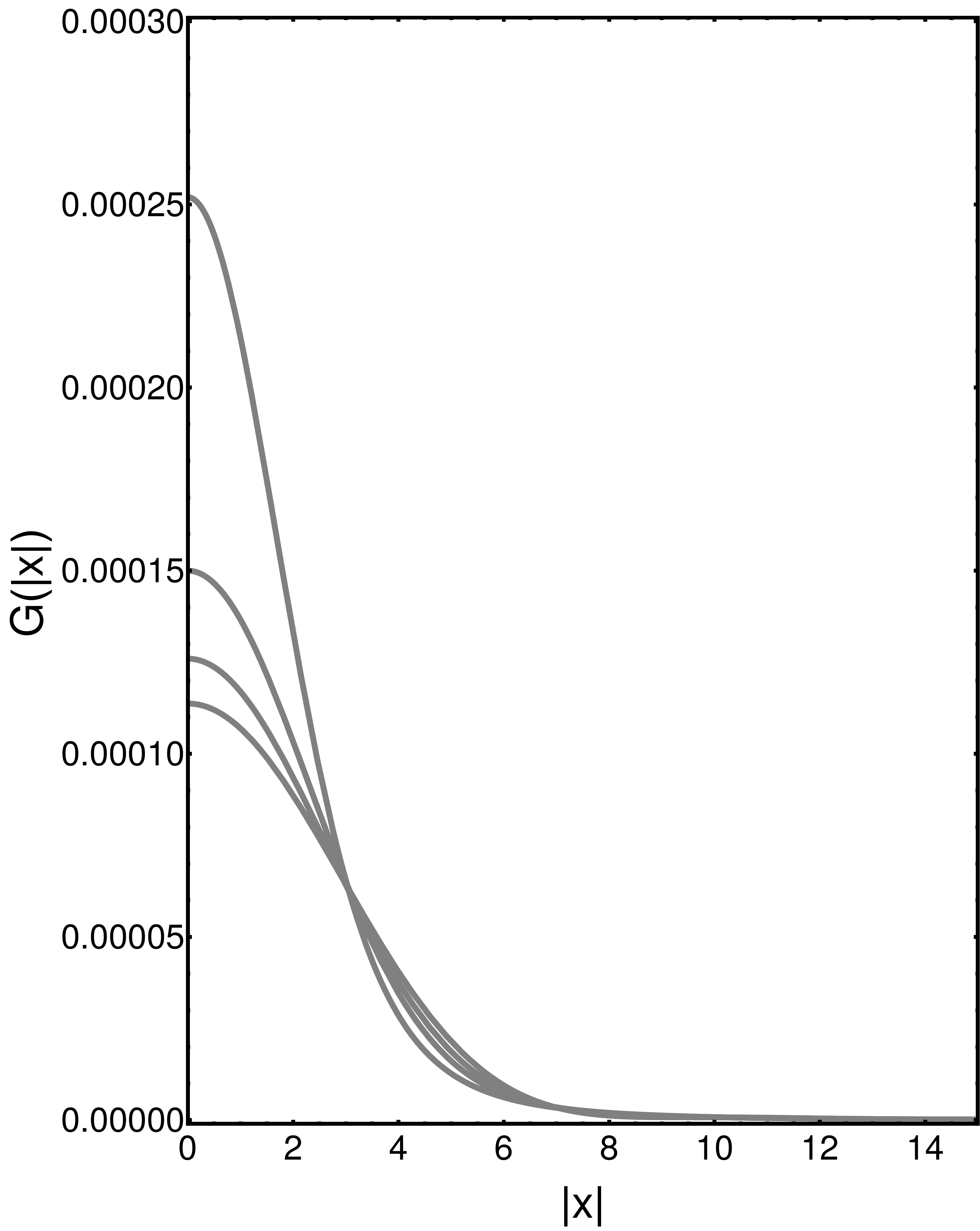}
\caption{The propagators for the form factor of eq. \eqref{expFF}} for $D=6$ and $n=1,3/2,2,3$. The behaviour is qualitatively similar as in Fig. \ref{fig:d4E}.
\label{fig:d6E}
\end{figure}

For completeness, we note that the propagators for the cases for which \eqref{eq:suff} does not hold can be studied numerically from eq.\! \eqref{propcoord}. Numerical studies did not reveal any example in which \eqref{RPcrit2} is violated, neither by considering higher dimensions nor by considering polynomials for which the sufficient condition \eqref{eq:suff} is violated. Indicative numerical results for all these cases are given 
in Figs.\,\ref{fig:d4E},\,\ref{fig:d4NL},\,\ref{fig:d4NLVio} and \ref{fig:d6E}.

\bigskip

{\em Conclusions ---} We gave a simple proof for the positivity of the propagator for a large class of nonlocal theories in four dimensions. As a consequence, a basic test for unitarity, perturbative reflection positivity in the coincidence limit, was shown to hold. This result covers theories defined by exponential form factors as well as form factors that have the properties required of candidates for a well defined quantum gauge or gravitational theory. 
Furthermore, we have provided numerical evidence that the same is true also for higher dimensions and for nonlocal theories not covered by the proof given here.  
These results contradict the claims in \cite{Asorey2} for the violation of perturbative reflection positivity in the coincidence limit for weakly nonlocal field theories, including quantum gravity or gauge theories. 

\pagebreak

\pagebreak
\end{document}